\documentclass[11pt,a4paper]{article}
\pdfoutput=1
\usepackage{jheppub}

\usepackage{color}
\usepackage{amsmath}
\usepackage{pifont}
\usepackage{bbold}

\usepackage{bbm}
\usepackage{verbatim}   
\usepackage{subfigure}
\usepackage{acronym}

\usepackage{amsfonts}
\usepackage{amssymb}
\usepackage{mathrsfs}
\usepackage{graphicx}
\usepackage{multirow}
 \usepackage{slashed}

\graphicspath{{./fig/}}

\definecolor{mypurple}{RGB}{164,64,214}

\def\bar#1{\overline{#1}}

\def\inv{^{\raise.15ex\hbox{${\scriptscriptstyle -}$}\kern-.05em 1}}
\def\lbar{{\lower.35ex\hbox{$\mathchar'26$}\mkern-10mu\lambda}} 

\let\<=\langle
\let\>=\rangle

\let\+=\uparrow

\newcommand{\newc}{\newcommand}
\newc{\gsim}{\lower.7ex\hbox{$\;\stackrel{\textstyle>}{\sim}\;$}}
\newc{\lsim}{\lower.7ex\hbox{$\;\stackrel{\textstyle<}{\sim}\;$}}

\title{LHCb anomalies from a natural perspective}

\author[a]{Isabel Garc\'ia Garc\'ia}
\emailAdd{isabel.garciagarcia@physics.ox.ac.uk}

\affiliation[a]{Rudolf Peierls Centre for Theoretical Physics,
University of Oxford,\\
1 Keble Road, Oxford,
OX1 3NP, UK}

\abstract{
Tension between the Standard Model (SM) and data concerning $b \rightarrow s$ processes has become apparent.
Most notoriously, concerning the $R_K$ ratio, which probes lepton non-universality in $b$ decays, and measurements involving the decays
$B \rightarrow K^* \mu^+ \mu^-$ and $B_s \rightarrow \phi \mu^+ \mu^-$.
Careful analysis of a wide range of $b \rightarrow s$ data shows that certain kinds of new physics can significantly ameliorate agreement with experiment.
Here, we show that these $b \rightarrow s$ anomalies can be naturally accommodated in the context of Natural Scherk-Schwarz Theories of the Weak Scale
-- a class of models designed to address the hierarchy problem.
No extra states beyond those naturally present in the theory need to be introduced in order to accommodate these anomalies,
and the assumptions required regarding flavor violating couplings are very mild.
Moreover, the structure of these models makes sharp predictions regarding $B$ meson decays into final states including $\tau^+ \tau^-$ pairs,
which will provide a future test of this type of theories.}

\begin{document}

\maketitle


\section{Introduction}
\label{sec:intro}

Hints of lepton non-universality in $b$-flavored meson decays have been reported by the LHCb experiment.
In particular, the ratio of the branching fractions $\mathcal{B}(B^+ \rightarrow K^+ \mu^+ \mu^-)$ to $\mathcal{B}(B^+ \rightarrow K^+ e^+ e^-)$
has been measured to be~\cite{Aaij:2014ora}:
\begin{equation}
	R_K \equiv \frac{\mathcal{B} (B^+ \rightarrow K^+ \mu^+ \mu^-)}{\mathcal{B} (B^+ \rightarrow K^+ e^+ e^-)} =
		0.745^{+ 0.090}_{-0.074} \ ({\rm stat}) \pm 0.036 \ ({\rm syst}) \ ,
\label{eq:RKmeasurement}
\end{equation}
which is $2.6 \sigma$ deviations away from the Standard Model (SM) prediction $R_K^{\rm SM} = 1.00 \pm 0.03$~\cite{Bordone:2016gaq}.
The measurement was performed for a dilepton invariant mass squared $q^2$ in the range $1 < q^2 / {\rm GeV}^2 < 6$,
and whereas $\mathcal{B} (B^+ \rightarrow K^+ e^+ e^-)$ seems to be consistent with the SM prediction,
the $\mathcal{B} (B^+ \rightarrow K^+ \mu^+ \mu^-)$ measurement falls below
\footnote{In the $1 < q^2 / {\rm GeV}^2 < 6$ region, LHCb has measured
	$\mathcal{B} ( B^+ \rightarrow K^+ e^+ e^- )_{[1,6]} = ( 1.56^{+0.19 \ +0.06}_{-0.15 \ -0.04} ) \cdot 10^{-7}$~\cite{Aaij:2014ora} and
	$\mathcal{B} ( B^+ \rightarrow K^+ \mu^+ \mu^- )_{[1,6]} = (1.19 \pm 0.03 \pm 0.06) \cdot 10^{-7}$~\cite{Aaij:2014pli},
	whereas the SM prediction for these branching fractions is $1.75^{+0.60}_{-0.29} \cdot 10^{-7}$~\cite{Bobeth:2012vn}.}.
This suggests that, if new physics (NP) is indeed behind the $R_K$ anomaly, it should affect muons stronger than electrons.

Although from a theoretical point of view $R_K$ is one of the cleanest observables involving $b \rightarrow s$ transitions that has been
observed to deviate from the SM, it is by no means the only one.
A plethora of observables in $b \rightarrow s$ decays have been measured, a number of which seem to be in mild disagreement with the SM --
most notably, the tension in $B \rightarrow K^* \mu^+ \mu^-$ angular observables~\cite{Aaij:2015oid},
and the branching fractions of the decays $B \rightarrow K^* \mu^+ \mu^-$~\cite{Aaij:2014pli} and $B^0_s \rightarrow \phi \mu^+ \mu^-$~\cite{Aaij:2013aln},
which seem to fall below their SM predictions~\cite{Ball:2004rg,Horgan:2013hoa,Horgan:2013pva,Straub:2015ica}.
A careful consideration of a large array of data involving $b \rightarrow s$ transitions was presented in~\cite{Altmannshofer:2014rta},
and NP contributions to certain Wilson coefficients seem to improve the fit to data compared to the SM.

In this article, we study the compatibility of these experimental anomalies with a well-motivated class of models that address the electroweak hierarchy problem,
those referred to as Natural Scherk-Schwarz Theories of the Weak Scale~\cite{Dimopoulos:2014aua,Garcia:2015sfa}.
This class of models solve the (little) hierarchy problem by combining supersymmetry (SUSY) and a flat extra spatial dimension, compactified on an $S^1/\mathbb{Z}_2$ orbifold and
with a compactification scale $1/R$ in the TeV range.
In the version of this theory described in~\cite{Dimopoulos:2014aua,Garcia:2015sfa}, all gauge, Higgs, and matter content are allowed to propagate in the extra-dimensional bulk,
except the third generation that remains localized on one of the 4-dimensional (4D) branes.
In the bulk, SUSY is broken non-locally by the Scherk-Schwarz mechanism (equivalent to breaking by boundary conditions).
The combination of the Scherk-Schwarz mechanism, and the localization of the third generation on one of the branes, allows this type of theories
to feature a very low level of fine-tuning, compared with standard 4D SUSY models~\cite{Arvanitaki:2013yja}.

\begin{figure}[h]
    \centering
    \includegraphics[scale=0.35]{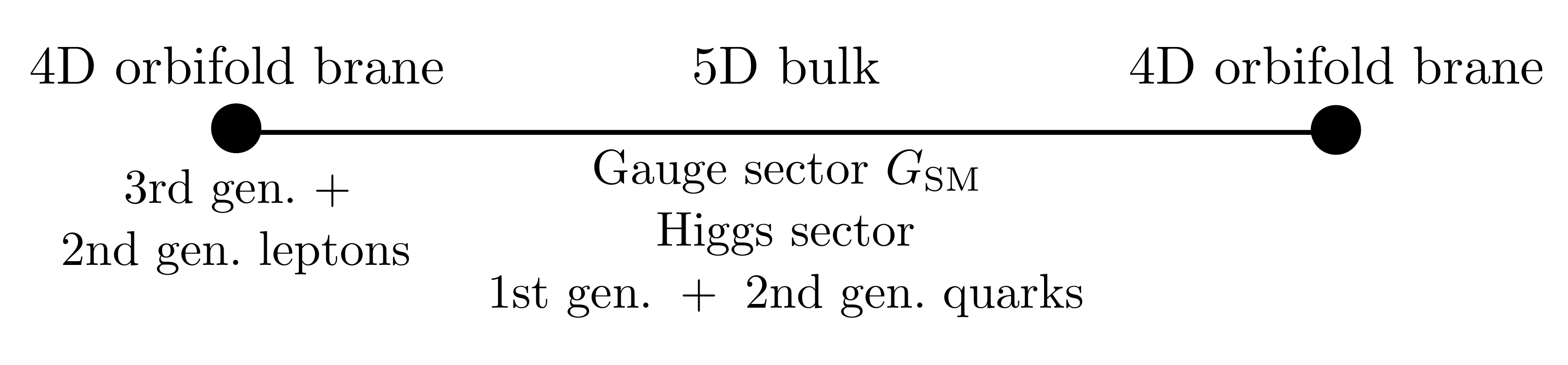}
\caption{Schematic illustration of the set-up considered in this work.
	Gauge and Higgs sectors propagate in the 5D bulk, together with the quarks of the second generation.
	The third generation, and the leptons of the second one, remain localized on one of the 4D branes.}
\label{fig:geography}
\end{figure}

In this work, we consider a small modification of the set-up described above, which consists in localizating the leptons of the second generation on one of the branes,
together with the third generation (see Fig.~\ref{fig:geography} for an illustration) \footnote{Effective 5D theories similar to those in~\cite{Dimopoulos:2014aua,Garcia:2015sfa}
have been previously considered in the
literature~\cite{Antoniadis:1998sd,Delgado:1998qr,Pomarol:1998sd,Barbieri:2000vh,Delgado:2001si,Delgado:2001xr,Barbieri:2002sw,Barbieri:2002uk,Marti:2002ar,Barbieri:2003kn,Diego:2005mu,Diego:2006py,vonGersdorff:2007kk,Bhattacharyya:2012ct,Larsen:2012rq},
although the models of~\cite{Dimopoulos:2014aua,Garcia:2015sfa} are the first of this kind
to be compatible with all current experimental constraints, and capable of accommodating a 125 GeV Higgs.}.
Although the original version of the theory is supersymmetric, we will see that the success of these models in accommodating the $R_K$ anomaly is completely independent
of SUSY, and only relies on the different localization of the matter content in the extra-dimension.
As a result of this different localization, the Kaluza-Klein (KK) excitations of the SM gauge bosons, the first of which appear at scale $1/R$,
will couple non-universally to fermions.
In particular, KK-modes of the $Z$ boson and the photon will couple maximally to $\mu^+ \mu^-$ (and $\tau^+ \tau^-$) pairs, whereas their couplings to $e^+ e^-$
pairs are naturally much more supressed.
This effect, combined with a certain amount of flavor violation present in the quark sector, allows this type of scenarios to accommodate
the observed deficit in the $R_K$ measurement, for values of $1/R$ in the $3-4~{\rm TeV}$ range.
Although we will focus on the $R_K$ measurement, we will see that the contributions to the Wilson coefficients we require are such that other $b \rightarrow s$ anomalies
are also ameliorated.

An interesting feature of this type of solutions to the $R_K$ anomaly is that other
$b$-flavored meson decays are sharply predicted to deviate from their SM predictions in a correlated manner.
In particular, the branching fractions for the processes $B^0_s \rightarrow \mu^+ \mu^-$, $B^0_s \rightarrow \tau^+ \tau^-$, and $B^+ \rightarrow K^+ \tau^+ \tau^-$
should all fall below the corresponding SM prediction by a factor $\sim R_K$.
As a matter of fact, $\mathcal{B}(B^0_s \rightarrow \mu^+ \mu^-)$ has been measured to differ from its SM expectation by a factor $0.77^{+0.20}_{-0.17}$
(see Appendix~\ref{app:Bsmumu} for details).
Although the experimental uncertainty is large, this measurement is certainly in the direction predicted by this kind of theories, and future,
more precise measurements will be a clean probe of the proposed set-up.
On the other hand, current experimental upper bounds on the branching fractions of the decays $B^0_s \rightarrow \tau^+ \tau^-$ and $B^+ \rightarrow K^+ \tau^+ \tau^-$
still lie around 4 orders of magntiude above the SM predictions (see Table~\ref{tab:BsDecays} and \ref{tab:BKdecays}).

Although the class of models considered in this work sharply predict non-universality between different generations of fermions, the extent to which extra
flavor violation is present remains largely arbitrary.
From a technical point of view, flavor violation (beyond that present in the SM) can be effectively turned-off by making the `right' choices, but under natural assumptions
some degree of flavor violation is predicted.
In the quark sector, its presence is a requisite in order to accommodate the $R_K$ anomaly, with $B^0_s - \overline{B^0}_s$ oscillation
measurements providing the leading constraint.
In the lepton sector, bounds on flavor violating decays allows us to gain insight into the flavor structure of these theories,
and we will see that the decays $\tau \rightarrow e \mu \mu$, $\tau \rightarrow 3 \mu$,
and $\mu \rightarrow 3e$ provide the most stringent limits.

Alternative explanations of these $b \rightarrow s$ anomalies already present in the literature include the presence of new spin-1 resonances
($Z'$)~\cite{Altmannshofer:2013foa,Gauld:2013qba,Crivellin:2015mga,Crivellin:2015lwa,Sierra:2015fma,Crivellin:2015era,Celis:2015ara,Falkowski:2015zwa,Descotes-Genon:2015uva,Carmona:2015ena,Allanach:2015gkd,Crivellin:2016ejn,Buttazzo:2016kid,Descotes-Genon:2016hem,Megias:2016bde},
or models including leptoquarks~\cite{Kosnik:2012dj,Hiller:2014yaa,Gripaios:2014tna,Sahoo:2015wya,Becirevic:2015asa,Becirevic:2016oho,Becirevic:2016yqi}.
Most of these scenarios need to assume the presence of extra states with the only purpose of accommodating anomalous $b \rightarrow s$ measurements.
An interesting exception is the work presented in~\cite{Megias:2016bde}, where a Randall-Sundrum set-up is considered and no extra states are assumed to
exist.
Like in~\cite{Megias:2016bde}, the class of models we consider are motivated by solving the electroweak hierarchy problem,
and an explanation of the $R_K$ anomaly is achieved as a result of interactions mediated by photon and $Z$ boson KK-modes,
which couple non-universally to muons and electrons due to the different profile of these fields along the extra dimension.
However, the flat nature of the extra dimension we consider leads to very different phenomenology.
In our case, no modification of the $Z$ boson couplings to fermions arises,
whereas in~\cite{Megias:2016bde} this is a major constraint that sets the lower bound on the first KK-mode mass.
For us, the lower bound on the compactification scale $1/R$ (and therefore on the mass of the first gauge boson KK-modes) is driven by gluino bounds,
which appear at scale $1/(2R)$ in this type of models.
Moreover, the fact that the third generation is also localized on the same brane as the muon sector leads to sharp predictions involving $B^+$ and $B^0_s$ decays
into $\tau$ final states that do not arise in~\cite{Megias:2016bde}.

The rest of the article is organised as follows.
In Section~\ref{sec:structure} we review the basic mechanism behind non-universal couplings in natural Scherk-Schwarz models.
Section~\ref{sec:RK} performs a detailed study of how the observed deficit in the $R_K$ measurement can be accommodated in this kind of theories,
and we study other correlated consequences.
We consider lepton flavor violating decays that might be induced in this context in Section~\ref{sec:FlavorViolatingDecays}, and derive bounds on the parameters of the model.
Our conclusions are presented in Section~\ref{sec:conclusions},
and a few appendices at the end contain the details of some of our calculations.

\section{Flavor Structure of Natural Scherk-Schwarz}
\label{sec:structure}

The presence of tree-level sources of flavor non-universality (and also flavor violation)
stems from the different localization of the different families in the extra dimension.
To illustrate this feature, we consider the couplings of leptons and down-type quarks to the photon field.
Starting from the 5D action, and after integrating over the extra dimension, we recover a 4D lagrangian that
explicitly shows how photon KK-modes couple non-universally to the different fermion families and, under natural assumptions,
also in a way that violates flavor.
The relevant terms in the 4D lagrangian read (using Dirac fermion notation)
\begin{equation}
\begin{aligned}
	\mathcal{L}_{\gamma, d} (x) &=
			\int_0^{\pi R} \ dy \ e_{\rm 5D} A_{\mu}^{\rm 5D} (x,y) Q_d \left\{
			\sum_{q=d,s} \bar q^{5D} (x,y) \gamma^\mu q^{5D} (x,y) +
			\delta(y) \ \bar b (x) \gamma^\mu b (x) \right\} \\
	\mathcal{L}_{\gamma, e} (x) &=
			\int_0^{\pi R} \ dy \ e_{\rm 5D} A_{\mu}^{\rm 5D} (x,y) Q_e \left\{
			\bar e^{5D} (x,y) \gamma^\mu e^{5D} (x,y) + \delta(y) \sum_{l=\mu,\tau} \bar l(x) \gamma^\mu l(x) \right\} \ ,
\end{aligned}
\label{eq:photon}
\end{equation}
for the couplings of the photon field to down-type quarks and charged leptons respectively.
Here, $e_{5D} = |e_{5D}|$ refers to the 5D electromagnetic coupling,
so that $e = e_{5D} / \sqrt{\pi R}$, and $Q_d = -1/3$, $Q_e = -1$.
The 5D fields $A_\mu^{5D}$ and $\psi^{5D}$ (for $\psi = e, d, s$) may be decomposed as
\begin{equation}
\begin{aligned}
	\phi^{5D} (x,y) &= \sum_{n=0}^{\infty} \phi^{5D (n)} (x) \cos \frac{n y}{R}  \\
				&= \frac{1}{\sqrt{\pi R}} \phi^{(0)}(x) + \sqrt{\frac{2}{\pi R}} \sum_{n=1}^{\infty} \phi^{(n)} (x) \cos \frac{n y}{R} \qquad (\phi=A_\mu, e, d, s) \ ,
\end{aligned}
\end{equation}
where $\phi^{(n)}$ corresponds to the $n$-th KK-mode of the corresponding field (appropriately 4D normalized),
and the 0-modes are to be identified with the corresponding SM particles.
Retaining only those terms in Eq.(\ref{eq:photon}) that involve fermion 0-modes, we find (suppressing the $x$ dependence)
\begin{equation}
\begin{aligned}
	\mathcal{L}^{(0)}_{\gamma, d} &= e Q_d A_\mu ( \bar q^d \gamma^\mu q^d )
			+ \sqrt2 e Q_d \left( \sum_{n=1}^{\infty} A_\mu^{(n)} \right) ( \bar q^d \gamma^\mu A_d q^d ) \\
	\mathcal{L}^{(0)}_{\gamma, e} &= e Q_e A_\mu ( \bar l^e \gamma^\mu l^e )
			+ \sqrt2 e Q_e \left( \sum_{n=1}^{\infty} A_\mu^{(n)} \right) ( \bar l^e \gamma^\mu A_e l^e ) \ ,
\end{aligned}
\label{eq:photon0}
\end{equation}
where $A_\mu \equiv A_\mu^{(0)}$, $q^d \equiv (d,  s,  b)^T$ and $l^e \equiv (e, \mu, \tau)^T$,
with $d \equiv d^{(0)}$, $s \equiv s^{(0)}$, and $e \equiv e^{(0)}$.
The matrices $A_d$ and $A_e$ encode the flavor structure of the photon $n \neq 0$ KK-mode interactions, and read
\begin{equation}
	A_d = \begin{pmatrix} 0 & 0 & 0 \\ 0 & 0 & 0 \\ 0 & 0 & 1 \end{pmatrix} \qquad {\rm and} \qquad A_e = \begin{pmatrix} 0 & 0 & 0 \\ 0 & 1 & 0 \\ 0 & 0 & 1 \end{pmatrix} .
\label{eq:Amatrix}
\end{equation}
Eq.(\ref{eq:photon0}) and (\ref{eq:Amatrix}) explicitly show how the 0-mode of the neutral gauge boson (in this example, the SM photon)
couples universally to all three generations, whereas heavier KK-modes do not.
In particular, in this gauge-eigenbasis the non-zero KK-modes
of gauge bosons couple to fermion fields in a way that is flavor diagonal but not flavor universal.

So far, we have worked within the flavor-eigenbasis of quarks and leptons.
However, to work with the physical fermion basis requires rotating the left-handed (LH) and right-handed (RH) matter fields by
$3 \times 3$ unitary matrices, e.g.~$q^d_J \rightarrow R^d_J q^d_J$  ($J=L,R$) for down-type quarks.
Under these rotations, interactions between the photon 0-mode and SM fermions remain unchanged,
but those involving higher KK-excitations are modified.
For instance, for down type quarks:
\begin{equation}
	\bar q^d \gamma^\mu A_d q^d \rightarrow \bar q^d_L \gamma^\mu (R^{d \dagger}_L A_d R^d_L) q^d_L + \bar q^d_R \gamma^\mu (R^{d \dagger}_R A_d R^d_R) q^d_R \ .
\end{equation}
Thus, if we define
\begin{equation}
	B^d_J \equiv R^{d \dagger}_J A_d R^d_J \qquad {\rm and} \qquad B^e_J \equiv R^{e \dagger}_J A_e R^e_J \qquad ({\rm for} \ J=L,R) \ ,
\label{eq:mixingMatrix}
\end{equation}
the couplings of photon KK-modes to fermion mass-eigenstates now read
\begin{equation}
\begin{aligned}
	\mathcal{L}^{(0)}_{\gamma, d} & = e Q_d A_\mu ( \bar q^d \gamma^\mu q^d )
			+ \sqrt2 e Q_d \left( \sum_{n=1}^{\infty} A_\mu^{(n)} \right) (\bar q^d_L \gamma^\mu B^d_L q^d_L + \bar q^d_R \gamma^\mu B^d_R q^d_R) \\
	\mathcal{L}^{(0)}_{\gamma, e} & = e Q_e A_\mu ( \bar l^e \gamma^\mu l^e )
			+ \sqrt2 e Q_e \left( \sum_{n=1}^{\infty} A_\mu^{(n)} \right) (\bar l^e_L \gamma^\mu B^e_L l^e_L + \bar l^e_R \gamma^\mu B^e_R l^e_R) \ .
\end{aligned}
\label{eq:finalLag}
\end{equation}
Notice that although the SM photon couples equally to LH and RH fermions, this is not true of its higher modes --
different rotation matrices for the LH and RH fermions will lead to different couplings.

At this point, it is illuminating to look at the structure of the $B^f_J$ matrices as a function of the entries of the different rotation matrices,
which can be written as (suppressing $J=L,R$ indices for clarity):
\begin{equation}
	B^d = \begin{pmatrix} |R_{31}|^2 & R_{31}^* R_{32} & R_{31}^* R_{33} \\ R_{32}^* R_{31} & |R_{32}|^2 & R_{32}^* R_{33} \\ R_{33}^* R_{31} & R_{33}^* R_{32} & |R_{33}|^2 \end{pmatrix}
	\sim \begin{pmatrix} \epsilon_d^2 & \epsilon_d^2 & \epsilon_d \\ \epsilon_d^2 & \epsilon_d^2 & \epsilon_d \\ \epsilon_d & \epsilon_d & 1 \end{pmatrix}
\label{eq:BmatrixQuarks}
\end{equation}
and
\begin{equation}
	B^e = \begin{pmatrix} 1-|R_{11}|^2 & -R_{11}^* R_{12} & -R_{11}^* R_{13} \\ -R_{12}^* R_{11} & 1-|R_{12}|^2 & R_{12}^* R_{13} \\ -R_{13}^* R_{11} & -R_{13}^* R_{12} & 1-|R_{13}|^2 \end{pmatrix}
	\sim \begin{pmatrix} \epsilon_e & \epsilon_e & \epsilon_e \\ \epsilon_e & 1 & \epsilon_e^2 \\ \epsilon_e & \epsilon_e^2 & 1 \end{pmatrix} ,
\label{eq:BmatrixLeptons}
\end{equation}
where in the last step we have assumed that the rotation matrices have a hierarchical structure, with diagonal terms being $\mathcal{O}(1)$ and off-diagonal terms
suppressed by a factor $\epsilon_d$ ($\epsilon_e$) for down-type quarks (leptons)
\footnote{The left-hand side of Eq.(\ref{eq:BmatrixQuarks}) is meant to merely illustrate how a suppression of the off-diagonal terms in the $B^d_J$ matrices
arises if the corresponding rotation matrices $R^d_J$ have a hierarchical structure.
The actual level of suppression between different flavor violating couplings in $B^d_L$ can only be specified in a given model where both $R^d_L$ and $R^u_L$
are known, and subject to the constraint $R^{u \dagger}_L R^d_L = V$ (with $V$ being the CKM matrix).
Such level of specification is unecessary to the present work.}.
This illustrates how, in this class of models, flavor non-universal effects are dominant compared to flavor violating ones.
At the same time, flavor violation between down-type quarks of the first and second generations is suppressed compared to flavor violation between the third generation
and the rest.
On the other hand, in the lepton sector, flavor violation between electrons and muons/taus dominates compared to that between muons and taus.
These two features are crucial in order to accommodate the $R_K$ anomaly, while remaining consistent with constraints from kaon mixing (see~\cite{Garcia:2014lfa} for a detailed study in this context) and the decay $\tau \rightarrow 3 \mu$
\footnote{In fact, it is possible to accommodate the $b \rightarrow s$ anomalies in scenarios where only the third generation is brane localized.
In such a set-up, however, the prediciton for $\mathcal{B} (\tau \rightarrow 3 \mu)$ is well above the current experimental upper bound.
Localizing the leptons of the second generation also on the brane alleviates this problem, as discussed in Sec.~\ref{sec:taumu}.}.

Although we have shown the appearance of tree-level non-universal, and flavor changing, neutral currents with the
example of the photon field and its couplings to down-type quarks and leptons, the same is true of all other neutral gauge bosons ($Z$ and gluons).
$W^\pm$ gauge boson KK-modes also lead to charged tree-level non-universal and flavor changing currents, but these will not be relevant for the present work.

Finally, we emphasize that the tree-level sources of flavor non-universality and flavor violation discussed in this section are the dominant contribution
to this type of processes, despite the underlying supersymmetric nature of the models we consider.
The reason why flavor changing interactions involving SUSY particles are subleading has to do with the absence of
a SUSY flavor problem in natural Scherk-Schwarz theories.
In particular, the fact that soft masses for the first and second generation squarks are dominantly flavor diagonal,
and the presence of a $U(1)_R$ symmetry that forbids $A$-terms and Majorana masses for gauginos and higgsinos,
vastly ameliorates the flavor situation compared to most 4D SUSY models.
For a detailed study of flavor physics (structure and constraints) in the context of natural Scherk-Schwarz scenarios, we refer the reader to~\cite{Garcia:2014lfa},
and to~\cite{Dimopoulos:2014aua,Garcia:2015sfa} for further details on the overall structure and phenomenology.

\section{$R_K$ and Correlated Effects}
\label{sec:RK}

\subsection{Formalism}
\label{sec:RKformalism}

The relevant set of effective operators for the study of the $B^+ \rightarrow K^+ l^+ l^-$ decay are
\begin{equation}
\begin{aligned}
	\mathcal{O}^{ll}_7 &= \frac{m_b}{e} (\bar b \sigma_{\mu \nu} P_L s) F^{\mu \nu} & \qquad \mathcal{O'}^{ll}_7 &= \frac{m_b}{e} (\bar b \sigma_{\mu \nu} P_R s) F^{\mu \nu} \\
	\mathcal{O}^{ll}_9 &= (\bar b \gamma^\nu P_L s) (\bar l \gamma_\nu l) & \qquad \mathcal{O'}^{ll}_9 &= (\bar b \gamma^\nu P_R s) (\bar l \gamma_\nu l) \\
	\mathcal{O}^{ll}_{10} &= (\bar b \gamma^\nu P_L s) (\bar l \gamma_\nu \gamma^5 l) & \qquad \mathcal{O'}^{ll}_{10} &= (\bar b \gamma^\nu P_R s) (\bar l \gamma_\nu \gamma^5 l) \ ,
\label{eq:operatorsRK}
\end{aligned}
\end{equation}
where $P_{L,R} = (1 \mp \gamma^5)/2$.
The effective hamiltonian can then be written as
\begin{equation}
	\mathcal{H}_{eff} = - \frac{4 G_F}{\sqrt{2}} \frac{\alpha}{4 \pi} (V_{tb}^* V_{ts}) \sum_{i=7,9,10} \left\{ C_i^{ll} \mathcal{O}^{ll}_i + {C'}_i^{ll} \mathcal{O'}^{ll}_i \right\} \ ,
\label{eq:Heff}
\end{equation}
where $C^{(') ll}_i = C^{(')}_{i, {\rm SM}} + C^{(') ll}_{i, {\rm NP}}$.
In the SM: $C_{9, {\rm SM}} \approx 4.23$, $C_{10, {\rm SM}} \approx -4.41$, $C_{7, {\rm SM}} \approx -0.32$~\cite{Khodjamirian:2010vf},
and $C^\prime_{7, {\rm SM}} = C^\prime_{9, {\rm SM}} = C^\prime_{10, {\rm SM}} = 0$.
Integrating out photon and $Z$ boson KK-modes, the following non-zero Wilson coefficients are generated
\begin{equation}
\begin{aligned}
	C^{\mu \mu}_{9, {\rm NP}} & \simeq - \frac{8 \pi^4}{3} \frac{v^2}{(1/R)^2} \frac{B^d_{L32}}{V^*_{tb} V_{ts}} ( S_{LL} B^e_{L22} + S_{LR} B^e_{R22} ) \\
	C^{' \mu \mu}_{9, {\rm NP}} & \simeq - \frac{8 \pi^4}{3} \frac{v^2}{(1/R)^2} \frac{B^d_{R32}}{V^*_{tb} V_{ts}} ( S_{RL} B^e_{L22} + S_{RR} B^e_{R22} ) \\
	C^{\mu \mu}_{10, {\rm NP}} & \simeq - \frac{8 \pi^4}{3} \frac{v^2}{(1/R)^2} \frac{B^d_{L32}}{V^*_{tb} V_{ts}} ( - S_{LL} B^e_{L22} + S_{LR} B^e_{R22} ) \\
	C^{' \mu \mu}_{10, {\rm NP}} & \simeq - \frac{8 \pi^4}{3} \frac{v^2}{(1/R)^2} \frac{B^d_{R32}}{V^*_{tb} V_{ts}} ( - S_{RL} B^e_{L22} + S_{RR} B^e_{R22} ) \ ,
\end{aligned}
\label{eq:WilsonCoeff}
\end{equation}
where $v \simeq 246 \ {\rm GeV}$, $S_{IJ} = 4 \alpha^d_I \alpha^e_J / \sin^2 2\theta_w + Q_d Q_e$ (for $I,J = L, R$),
and $\alpha^f_L = -1/2 - Q_f \sin^2 \theta_w$, $\alpha^f_R = - Q_f \sin^2 \theta_w$, with $Q_d = -1/3$ and $Q_e=-1$.
\footnote{Numerically: $S_{LL} \simeq 0.97$, $S_{LR} \simeq -0.22$, $S_{RL} \simeq 0.22$ and $S_{RR} \simeq 0.43$.}
(Wilson coefficients relevant to the operators involving electrons rather than muons are the same as those in Eq.(\ref{eq:WilsonCoeff})
after performing the obvious substitution $B^e_{J 22} \rightarrow B^e_{J 11}$.)
Notice no $C^{(\prime) \mu \mu}_{7, {\rm NP}}$ are generated, and since $|C_{7, {\rm SM}} / C_{9, {\rm SM}}| \sim 0.07$, we will neglect
the effect of $C_{7, {\rm SM}}$ in the following.
The prediction for $R_K$ is then approximately given by
\begin{equation}
	R_K \simeq \frac{ | C_{9, {\rm SM}} + C^{\mu \mu}_{9, {\rm NP}} + C^{' \mu \mu}_{9, {\rm NP}} |^2 +
	| C_{10, {\rm SM}} + C^{\mu \mu}_{10, {\rm NP}} + C^{' \mu \mu}_{10, {\rm NP}} |^2}{| C_{9, {\rm SM}} + C^{ee}_{9, {\rm NP}} + C^{' ee}_{9, {\rm NP}} |^2 +
	| C_{10, {\rm SM}} + C^{ee}_{10, {\rm NP}} + C^{' ee}_{10, {\rm NP}} |^2}
\label{eq:RK}
\end{equation}

Theoretical analyses regarding $b \rightarrow s l^+ l^-$ processes (including the $R_K$ measurement discussed here) suggest that the best fit
to data is achieved if new non-zero contributions are present for those Wilson coefficients that involve muons rather than electrons,
and those scenarios in which only $C^{\mu \mu}_{9, {\rm NP}} \neq 0$,  only $C^{\mu \mu}_{10, {\rm NP}} \neq 0$,
or $C^{\mu \mu}_{9, {\rm NP}} = - C^{\mu \mu}_{10, {\rm NP}} \neq 0$ seem to be preferred~\cite{Altmannshofer:2014rta,Ghosh:2014awa}.
In our set-up, Wilson coefficients of the effective operators involving electrons are expected to be much smaller than those involving muons,
due to the natural supression in the $B^e_{J11}$ couplings compared to $B^e_{J22} \approx 1$ (see Eq.(\ref{eq:BmatrixLeptons})),
and so we neglect the effects of the former in what follows.
Realizing a scenario in which only one of the Wilson coefficients is non-zero is not possible in our set-up (in the absence of barroque arrangements),
but the scenario where $C^{\mu \mu}_{9, {\rm NP}} \sim - C^{\mu \mu}_{10, {\rm NP}}$ is naturally realized taking into account that one
expects $B^e_{J22} \approx 1$ (for both $J=L,R$), and observing the hierarchy $|S_{LR}|/|S_{LL}| \sim 0.2$.
For simplicity in performing our analysis, in this work we will make the assumption that $|B^d_{R32}| \ll |B^d_{L32}|$,
which allows us to neglect the coefficients $C^{' \mu \mu}_{i, {\rm NP}}$ in favor of $C^{\mu \mu}_{i, {\rm NP}}$ ($i=9,10$)
\footnote{Although the assumption $|B^d_{R32}| \ll |B^d_{L32}|$ appears unjustified,
we remind the reader that the aim of this work is not to achieve a full explanation of the LHCb anomalies,
but rather to assess whether it is possible to accommodate them in the framework of the class of theories described in~\cite{Dimopoulos:2014aua,Garcia:2015sfa}.
A real explanation of this assumption would require a full model that specifies the structure of RH and LH rotation matrices in the quark sector,
something which is beyond the scope of this work, albeit we note that allowing $b_R$ to propagate in the bulk would achieve such suppression.}.
Only three parameters are then left in order to accommodate the $R_K$ anomaly: the compactification scale $1/R$, and the magnitude and phase of the coupling $B^d_{L32}$.
We find it is convenient to parametrise $B^d_{L32}$ as
\begin{equation}
B^d_{L32} \equiv V^*_{tb} V_{ts} \frac{|B^d_{L32}|}{|V^*_{tb} V_{ts}|} e^{i \delta} \ ,
\label{eq:BdL32}
\end{equation}
such that $\delta \neq 0$ corresponds to a non-zero phase relative to the SM amplitude for the $b \rightarrow s l^+ l^-$ process
that underlies the $B^+ \rightarrow K^+ l^+ l^-$ decay.
The relevant Wilson coefficients therefore simplify to
\begin{equation}
\begin{aligned}
	C^{\mu \mu}_{9, {\rm NP}} & \simeq - 0.7 \left( \frac{3 \ {\rm TeV}}{1/R} \right)^2 \frac{| B^d_{L32} |}{0.5 \cdot |V^*_{tb} V_{ts}|} e^{i \delta} \qquad
	& C^{' \mu \mu}_{9, {\rm NP}} & \simeq 0 \\
	C^{\mu \mu}_{10, {\rm NP}} & \simeq + 1.0 \left( \frac{3 \ {\rm TeV}}{1/R} \right)^2 \frac{| B^d_{L32} |}{0.5 \cdot |V^*_{tb} V_{ts}|} e^{i \delta} \qquad
	& C^{' \mu \mu}_{10, {\rm NP}} & \simeq 0 \ ,
\end{aligned}
\label{eq:WilsonCoeffv2}
\end{equation}
and Eq.(\ref{eq:RK}) then reduces to
\begin{equation}
	R_K \simeq \frac{ | C_{9, {\rm SM}} + C^{\mu \mu}_{9, {\rm NP}} |^2 + | C_{10, {\rm SM}} + C^{\mu \mu}_{10, {\rm NP}} |^2}{|C_{9, {\rm SM}}|^2 + |C_{10, {\rm SM}}|^2}
\label{eq:RKv2}
\end{equation}

Since $| C_{i, {\rm SM}} + C^{\mu \mu}_{i, {\rm NP}} |^2 = |C_{i, {\rm SM}}|^2 + |C^{\mu \mu}_{i, {\rm NP}}|^2 - 2 |C_{i, {\rm SM}}| |C^{\mu \mu}_{i, {\rm NP}}| \cos \delta$ (for $i=9,10$),
it is clear that a decrease in $R_K$ compared to the SM prediction can only be achieved for $\delta \subset (-\pi/2, \pi/2)$,
and so we only consider values of $\delta$ within this range in what follows.

\subsection{Results}
\label{sec:RKresults}

In the context of natural Scherk-Schwarz theories, a low level of fine-tuning in the electroweak sector is achieved
for a compactification scale $1/R$ of a few TeV.
Lower bounds on $1/R$ stem from different sources, but mainly from gluino searches (gluinos are predicted to have masses of size $1/(2R)$),
and $Z'$ searches (in our case, the lowest lying $Z'$ would be the first KK-mode of the $Z$ gauge boson).
Regarding the former, the unsual spectrum of this kind of theories makes it difficult to translate current experimental bounds to our particular scenario
(see~\cite{Dimopoulos:2014aua,Garcia:2015sfa} for details on the spectrum and phenomenology of these models).
However, current bounds suggest that a gluino mass $m_{\tilde g} \approx 1.5 \ {\rm TeV}$ is very likely to be allowed in our case,
setting a conservative lower bound on $1/R$ of around $3 \ {\rm TeV}$
\footnote{Part of the reason why standard experimental constraints on stops and gluino masses do not straightforwardly apply to natural Scherk-Schwarz
models is that stops feature a three-body decay process with several invisible particles in the final state, which is known to significantly weaken constraints~\cite{Alves:2013wra}.
Gluinos, being much heavier than stops, typically decay to a top-stop pair, the stop then decaying.
Moreover, a compressed sparticle spectrum is typically expected in the context of natural Scherk-Schwarz,
which would further weaken experimental limits~\cite{Dimopoulos:2014psa}.}.

Regarding $Z'$ searches, bounds are very stringent if the couplings of the $Z'$ are the same as those of the $Z$ --
according to~\cite{ATLAS:2016cyf}, the lower bound on the $Z'$ mass may be as high as $\sim 4 \ {\rm TeV}$.
In our case, couplings of the lightest $Z$ KK-mode to first and second generation quarks are constrained to be very suppressed due
to kaon mixing measurements (see \cite{Garcia:2014lfa}, where this issue was discussed in detail),
and the production of this $Z'$ state in proton-proton collisions is largely due to annihilation of $\bar b b$ pairs.
However, for proton collisions at the centre of mass energy relevant for the LHC (i.e.~$\sqrt{s} \approx 14 \ {\rm TeV}$),
and for a parton momentum fraction $x \approx 0.2$ (the typical value for production of a $3-4 \ {\rm TeV}$ resonance),
the parton-parton luminosities of $\bar b b$ pairs compared to that of light quarks are smaller by a factor of $10^{-3}$
(see e.g.~Figure~8.(a) in \cite{PDF}).
As a result, even though the coupling of the lightest KK-mode to $\bar b b$ pairs is enhanced
by a factor of $\sqrt{2}$, the total production cross section compared to that of a $Z'$ with the same couplings as those of the SM $Z$
is smaller by a factor $\sim 10^{-3}$.
Thus, even allowing for an $\mathcal{O}(1)$ increase in $\mathcal{B} (Z' \rightarrow \mu^+ \mu-)$ compared to the SM case,
the decrease in the production cross section is so large that a $Z'$ as light as $1.5 \ {\rm TeV}$ seems consistent with current limits --
well below the range of masses that we consider.

\begin{figure}[h]
    \centering
    \includegraphics[scale=0.84]{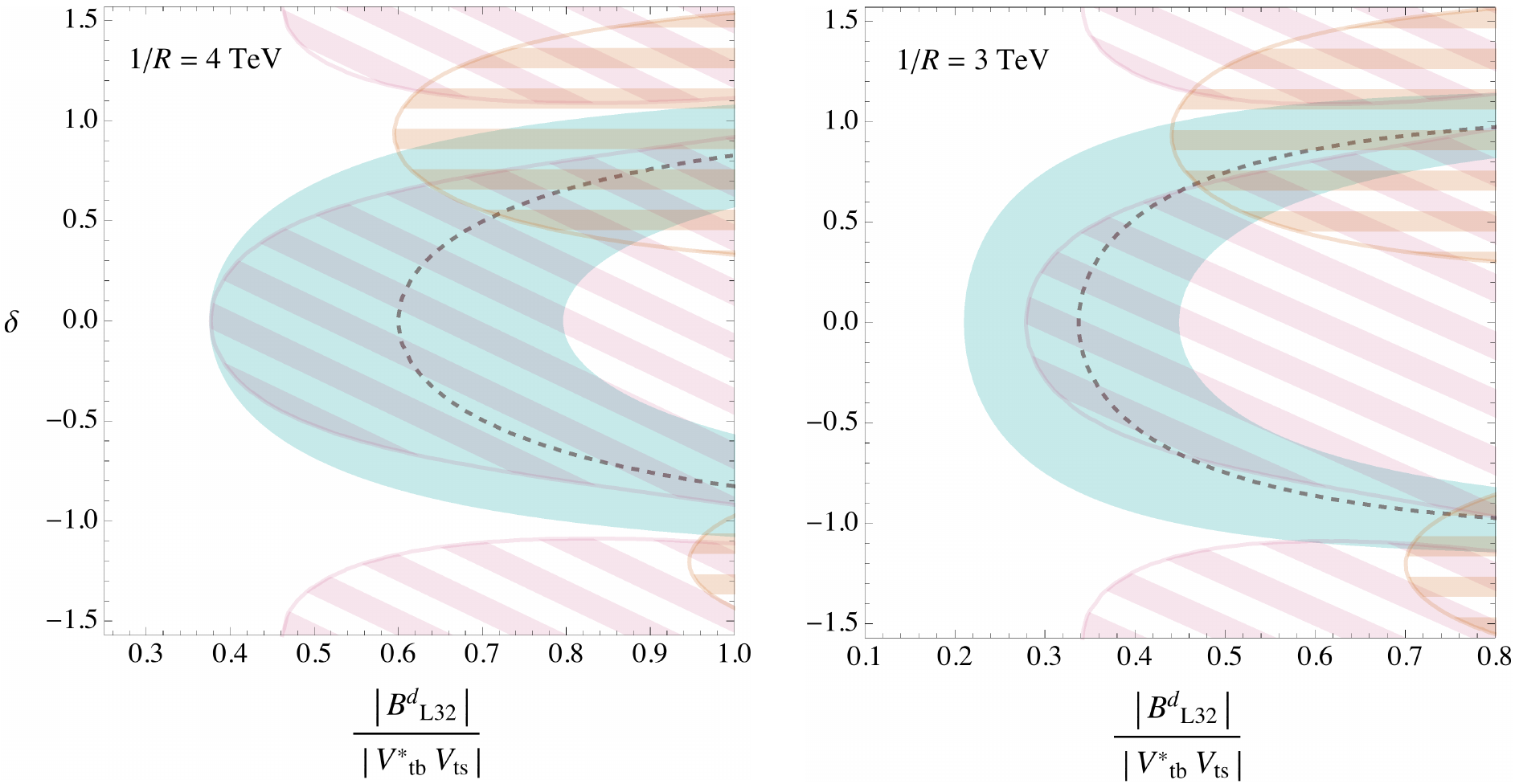}
\caption{The blue area corresponds to the region of parameter space where the predicted value of $R_K$ falls within the $1 \sigma$ region
	allowed by the LHCb measurement, with the dashed gray line corresponding to the central value $R_K = 0.745$.
	The area shaded with pink diagonal lines is ruled out by $\Delta M_s$ measurements,
	whereas the area shaded with orange horizontal lines is ruled out by measurements on the CP asymmetry in the $B^0_s - {\overline B}^0_s$ system
	(see Appendix~\ref{app:Bsmixing} for details).
	The compactification scale is $1/R = 4 \ {\rm TeV}$ and $3 \ {\rm TeV}$ in the left and right plots respectively.}
\label{fig:contour}
\end{figure}

\begin{figure}[h]
    \centering
    \includegraphics[scale=0.84]{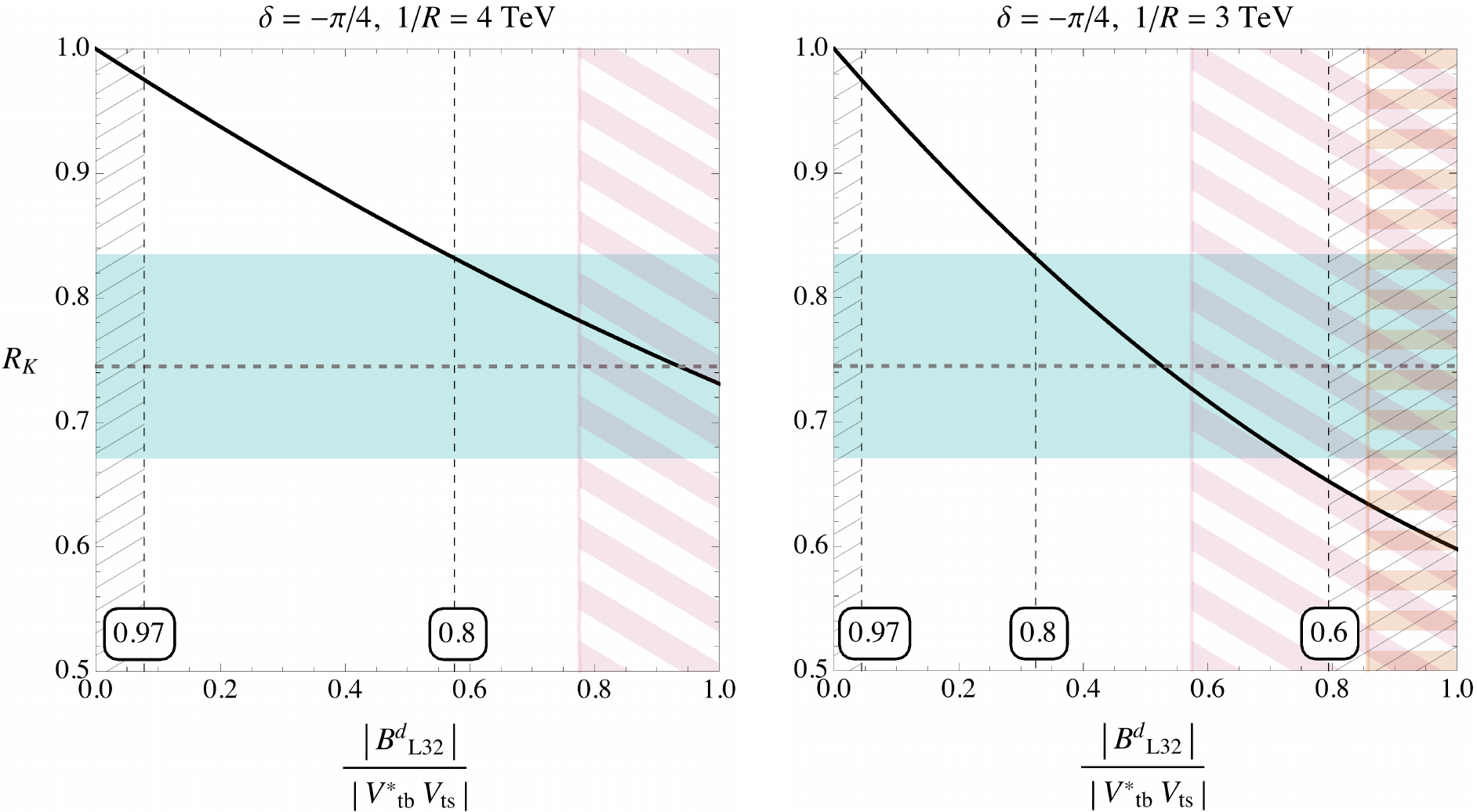}
\caption{The continuous black line represents the predicted value of $R_K$ as a function of $|B^d_{L,32}|$ for $\delta = -\pi/4$.
	The blue area corresponds to those values of $R_K$ within the $1 \sigma$ region allowed by the LHCb measurement,
	with the horizontal gray dashed line corresponding to the central value $R_K = 0.745$.
	Vertical dashed lines correspond to constant values of the ratio $\mathcal{B} (B^0_s \rightarrow \mu^+ \mu^-)$ to its prediction in the SM, as indicated by the numbers,
	and the regions dashed with thin gray lines fall beyond the $1\sigma$ band allowed for this ratio (see Appendix~\ref{app:Bsmumu} for details).
	Regions dashed with pink and orange lines are ruled out by measurements on the $B^0_s - {\overline B}^0_s$ system,
	as described in the caption of Fig.~\ref{fig:contour}.
	The compactification scale is $1/R = 4 \ {\rm TeV}$ and $3 \ {\rm TeV}$ in the left and right plots respectively.}
\label{fig:deltamPi4}
\end{figure}

Once we choose $1/R$ to be in the TeV scale, the range motivated by naturalness,
the only two parameters left to accommodate the $R_K$ anomaly are the magnitude of the $B^d_{L32}$ coupling,
and the relative phase $\delta = {\rm arg}(B^d_{L32}) - {\rm arg}(V^*_{tb} V_{ts})$.
In Figure~\ref{fig:contour} we show the region of parameter space (in the $|B^d_{L32}| - \delta$ plane) that allows the measured $R_K$ deficit to be accommodated,
for $1/R = 3, 4 \ {\rm TeV}$.
The main constraints on both $|B^d_{L32}|$ and $\delta$ arise from measurements regarding $B^0_s - {\overline B}^0_s$ oscillations,
represented in the figure by the pink and orange dashed regions respectively (see Appendix~\ref{app:Bsmixing} for details regarding these constraints).
As can be appreciated, some region of parameter space consistent with $B^0_s - {\overline B}^0_s$ measurements remains that can accommodate
the $R_K$ anomaly, in particular in the $1/R = 3 \ {\rm TeV}$ case, with the $1/R = 4 \ {\rm TeV}$ scenario only allowing marginal agreement at the $1\sigma$ level.
Figure~\ref{fig:deltamPi4} shows the predicted value of $R_K$ for $1/R = 3, 4 \ {\rm TeV}$ for the particular choice $\delta = - \pi/4$,
for which the constraints on $|B^d_{L32}|$ are weakest.

The NP contributions to the Wilson coefficients of Eq.(\ref{eq:WilsonCoeff}) have further consequences than simply altering the $R_K$ ratio,
and it is crucial to notice that the sizes of $C^{\mu \mu}_{9, {\rm NP}}$ and $C^{\mu \mu}_{10, {\rm NP}}$ that we require to accommodate the $R_K$ anomaly
are also in the region preferred by other $b \rightarrow s$ measurements~\cite{Altmannshofer:2014rta,Ghosh:2014awa}.
For instance, a second process described by the same set of effective operators is the decay $B^0_s \rightarrow \mu^+ \mu^-$,
whose branching fraction is predicted to differ from its SM value by a factor $R^s_{\mu \mu}$ given by
\begin{equation}
	R^s_{\mu \mu} \equiv \frac{ \mathcal{B} (B^0_s \rightarrow \mu^+ \mu^-) }{ \mathcal{B} (B^0_s \rightarrow \mu^+ \mu^-)_{\rm SM} }
		\simeq \frac{ |C_{10, {\rm SM}} + C^{\mu \mu}_{10, {\rm NP}} - C^{\prime \mu \mu}_{10, {\rm NP}}|^2 }{|C_{10, {\rm SM}}|^2}
		\simeq \frac{ |C_{10, {\rm SM}} + C^{\mu \mu}_{10, {\rm NP}}|^2}{|C_{10, {\rm SM}}|^2} \ ,
\label{eq:Rsmumu}
\end{equation}
where in the last step we have neglected the contribution from $C^{\prime \mu \mu}_{10, {\rm NP}}$, as discussed in Section~\ref{sec:RKformalism}.
Given that $C_{10, {\rm SM}} \sim - C_{9, {\rm SM}}$, and $C^{\mu \mu}_{10, {\rm NP}} \sim - C^{\mu \mu}_{9, {\rm NP}}$, it becomes obvious by comparing
Eq.(\ref{eq:Rsmumu}) and Eq.(\ref{eq:RKv2}) that $R^s_{\mu \mu} \sim R_K$: accommodating the $R_K$ anomaly necessarily implies that the value of
$\mathcal{B} (B^0_s \rightarrow \mu^+ \mu^-)$ should deviate from its SM prediction by a factor $R^s_{\mu \mu}$ which is of the same size as $R_K$.
The current exprimental measurement of $\mathcal{B} (B^0_s \rightarrow \mu^+ \mu^-)$ constrains this ratio to be $R^s_{\mu \mu} = 0.77^{+0.20}_{-0.17}$
(see Appendix~\ref{app:Bsmumu} for details), which certainly agrees with our prediction $R^s_{\mu \mu} \sim R_K$, although admittedly the experimental uncertainty in the
$\mathcal{B} (B^0_s \rightarrow \mu^+ \mu^-)$ measurement is currently too large for strong statements to be made.
A more precise determination of the $B^0_s \rightarrow \mu^+ \mu^-$ branching fraction, and its correlation with the measured value of $R_K$,
will provide a test of these scenarios.

Regarding $\tau^+ \tau^-$ final states, the decays $B^0_s \rightarrow \tau^+ \tau^-$ and $B^+ \rightarrow K^+ \tau^+ \tau^-$ are described by the same operators of Eq.(\ref{eq:operatorsRK})
after performing the obvious substitution $\mu \rightarrow \tau$.
The NP contributions to their Wilson coefficients, $C^{(\prime) \tau \tau}_{i, {\rm NP}}$ ($i=9,10$), are those of Eq.(\ref{eq:WilsonCoeff}) after substituting $B^e_{J22} \rightarrow B^e_{J33}$.
However, the structure of the theory is such that $B^e_{J22} \simeq B^e_{J33} \simeq 1$ (see Eq.(\ref{eq:BmatrixLeptons})), which implies
$C^{\tau \tau}_{i, {\rm NP}} \simeq C^{\mu \mu}_{i, {\rm NP}}$, and $C^{\prime \tau \tau}_{i, {\rm NP}} \simeq C^{\prime \mu \mu}_{i, {\rm NP}}$ ($i=9,10$).
As a result, the branching fractions for the decays $B^0_s \rightarrow \tau^+ \tau^-$ and $B^+ \rightarrow K^+ \tau^+ \tau^-$ are also predicted to
deviate from their SM values by a factor similar to $R_K$.
At the moment, only an upper bound on the branching fraction of these two decays exist, which is of $\mathcal{O}(10^{-3})$~\cite{BstautauExp,TheBaBar:2016xwe},
whereas the SM predicts a branching fraction of $\mathcal{O}(10^{-6})$ and $\mathcal{O}(10^{-7})$
for $B^0_s \rightarrow \tau^+ \tau^-$ and $B^+ \rightarrow K^+ \tau^+ \tau^-$ respectively~\cite{Bobeth:2013uxa,Bouchard:2013mia}
(see Table~\ref{tab:BsDecays} and \ref{tab:BKdecays}).
No information can therefore be extracted from these two decay modes at the moment,
but future measurements will be a key probe of these theories.

Finally, regarding the $B^0_s \rightarrow e^+ e^-$ decay channel, due to the naturally small coupling between $Z$ and photon KK-modes to first generation leptons,
no significant deviation of this branching fraction compared to the SM prediction is expected,
in the same way that $\mathcal{B} (B^+ \rightarrow K^+ e^+ e^-) \simeq \mathcal{B} (B^+ \rightarrow K^+ e^+ e^-)_{\rm SM}$ is predicted in these models.
Only a weak upper bound currently exist for $\mathcal{B} (B^0_s \rightarrow e^+ e^-)$ of $\mathcal{O}(10^{-7})$~\cite{Olive:2016xmw} --
roughly six orders of magnitude above the SM prediction~\cite{Bobeth:2013uxa} (see Table~\ref{tab:BsDecays}).

\section{Lepton Flavor Violating Decays}
\label{sec:FlavorViolatingDecays}

Lepton flavor violating couplings of the $Z$ and photon KK-modes are encoded in the off-diagonal elements of the $B^e_J$ matrices.
Given the structure of the models considered in this work (different localization of the first generation leptons compared to those of the second and third families),
the processes that can best constrain these flavor violating couplings are the decays
$\tau \rightarrow e \mu \mu$, $\tau \rightarrow 3 \mu$, and $\mu \rightarrow 3 e$
\footnote{Details regarding the exact expressions for these branching fractions are contained in Appendix~\ref{app:LeptonDecays}.}.
In this section, we discuss how current upper bounds affect the flavor structure of natural Scherk-Schwarz models in the lepton sector.

\subsection{$\tau \rightarrow e$}
\label{sec:taue}
Regarding the $\tau \rightarrow e \mu \mu$ decay, we find
\begin{equation}
	\mathcal{B} (\tau^- \rightarrow e^- \mu \mu) \approx 4 \cdot 10^{-6}  \left( \frac{3 \ {\rm TeV}}{1/R} \right)^4 |B^e_{13}|^2 \ ,
\end{equation}
where in the last step we have used the fact that $|B^e_{J22}| \simeq 1$ (for $J=L,R$), and we have set $B^e_{L13} = B^e_{R13} \equiv B^e_{13}$ for simplicity.
The current experimental upper bound on this decay is $\mathcal{B} (\tau^- \rightarrow e^- \mu \mu) < 2.7 \cdot 10^{-8}$~\cite{Hayasaka:2010np},
which imposes a mild constraint $|B^e_{13}| \lesssim 0.08 \ (0.14)$ for $1/R = 3 \ (4) \ {\rm TeV}$.
Given that $|B^e_{13}| = |R^e_{11}| |R^e_{13}| \approx |R^e_{13}|$ for $|R^e_{11}| \approx 1$,
this translates directly into a constraint on $|R^e_{13}|$ of the same size.

With this upper bound on the size of $|B^e_{J13}|=|B^e_{J31}|$, and the values of $|B^d_{L32}|$ necessary to fit the $R_K$ and other $b \rightarrow s$ anomalies,
an upper bound on the predicted branching fraction of two other processes can be computed:
the decays $B^0_s \rightarrow \tau^\mp e^\pm$ and $B^+ \rightarrow K^+ \tau^\mp e^\pm$.
These are given by
\begin{equation}
	\mathcal{B} (B^0_s \rightarrow \tau^\mp e^\pm) \approx 2 \cdot 10^{-10}
					\left( \frac{3 \ {\rm TeV}}{1/R} \right)^4		\left( \frac{|B^e_{31}|}{0.08} \right)^2	\left( \frac{|B^d_{L32}|}{0.5 |V^*_{tb} V_{ts}|} \right)^2
\end{equation}
\begin{equation}
	\mathcal{B} (B^+ \rightarrow K^+ \tau^\mp e^\pm) \approx 5 \cdot 10^{-11}
					\left( \frac{3 \ {\rm TeV}}{1/R} \right)^4		\left( \frac{|B^e_{31}|}{0.08} \right)^2	\left( \frac{|B^d_{L32}|}{0.5 |V^*_{tb} V_{ts}|} \right)^2 \ .
\end{equation}
Both branching fractions lie very far below current experimental upper bounds (see Table~\ref{tab:BsDecays} and \ref{tab:BKdecays}).

\subsection{$\tau \rightarrow \mu$}
\label{sec:taumu}

Regarding the decay channel $\tau \rightarrow 3 \mu$, we find the following branching fraction
\begin{equation}
	\mathcal{B} (\tau \rightarrow 3 \mu) \approx 1 \cdot 10^{-5}  \left( \frac{3 \ {\rm TeV}}{1/R} \right)^4 |B^e_{23}|^2 \ ,
\end{equation}
where we have taken into account the fact that $|B^e_{J22}| \simeq 1$ (for $J=L,R$), and we have set $B^e_{L23} = B^e_{R23} \equiv B^e_{23}$ for simplicity.
The current experimental upper bound on this decay is $\mathcal{B} (\tau \rightarrow 3\mu) < 2.1 \cdot 10^{-8}$~\cite{Hayasaka:2010np},
which imposes a constraint $|B^e_{23}| = |B^e_{32}| \lesssim 0.04 \ (0.07)$ for $1/R = 3 \ (4) \ {\rm TeV}$.
Given that $|B^e_{23}| = |R^e_{12}| |R^e_{13}|$, if $|R^e_{12}| \sim |R^e_{13}|$ then the constraint on $|B^e_{23}|$ translates into a mild upper bound
$|R^e_{12}|, |R^e_{13}| \lesssim 0.20 \ (0.26) $.

With this upper limit on the size of $|B^e_{23}| = |B^e_{32}|$, and the values of $|B^d_{L32}|$ necessary to fit the $R_K$ anomaly,
an upper bound on the predicted branching fraction for
the decays $B^0_s \rightarrow \tau^\mp \mu^\pm$ and $B^+ \rightarrow K^+ \tau^\mp \mu^\pm$ can be set:
\begin{equation}
	\mathcal{B} (B^0_s \rightarrow \tau^\mp \mu^\pm) \approx 6 \cdot 10^{-11}
					\left( \frac{3 \ {\rm TeV}}{1/R} \right)^4		\left( \frac{|B^e_{32}|}{0.04} \right)^2	\left( \frac{|B^d_{L32}|}{0.5 |V^*_{tb} V_{ts}|} \right)^2
\end{equation}
\begin{equation}
	\mathcal{B} (B^+ \rightarrow K^+ \tau^\mp \mu^\pm) \approx 1 \cdot 10^{-11}
					\left( \frac{3 \ {\rm TeV}}{1/R} \right)^4		\left( \frac{|B^e_{32}|}{0.04} \right)^2	\left( \frac{|B^d_{L32}|}{0.5 |V^*_{tb} V_{ts}|} \right)^2 \ .
\end{equation}
Again, both values are well below current constraints (see Table~\ref{tab:BsDecays} and \ref{tab:BKdecays}).

\subsection{$\mu \rightarrow e$}
\label{sec:mue}

Finally, we consider the decay $\mu \rightarrow 3e$, whose branching fraction we find to be
\begin{equation}
	\mathcal{B} (\mu \rightarrow 3 e) \approx 7.8 \cdot 10^{-13} \left( \frac{3 \ {\rm TeV}}{1/R} \right)^4 \left( \frac{|B^e_{11} B^e_{12}|}{10^{-4}} \right)^2 \ ,
\end{equation}
again assuming $B^e_{L11} = B^e_{R11} \equiv B^e_{11}$, and $B^e_{L12} = B^e_{R12} \equiv B^e_{12}$ for simplicity.
The very stringent experimental upper bound on this branching fraction, $\sim 1.0 \cdot 10^{-12}$~\cite{Bellgardt:1987du},
imposes a constraint $|B^e_{11} B^e_{12}| \lesssim 10^{-4} \ (2 \cdot 10^{-4})$ for $1/R = 3, \ (4)$ TeV.
Taking into account that $B^e_{11} = 1 - |R^e_{11}|^2$, $|B^e_{12}| = |R^e_{11}| |R^e_{12}|$, and that we expect $R^e_{11} \approx 1$,
constraints are satisfied, for instance, for $|R^e_{11}| \approx 0.99$ and $|R^e_{12}| \approx 0.005$,
so that $|B^e_{11}| \approx 0.02$ and $|B^e_{12}| \approx 0.005$.
The fact that, as we take $|R^e_{11}| \rightarrow 1$, the prediction for $\mathcal{B} (\mu \rightarrow 3 e)$ vanishes,
makes it imposible to put an absolute bound on the size of $B^e_{12}$.
However, it is clear that this process imposes non-trivial constraints on the flavor structure of these models, and future measurements performed
by the Mu3e experiment will shed even more light in this direction, since their expected sensitivity is as low as $10^{-16}$~\cite{Blondel:2013ia}.

Other processes involving $\mu - e$ flavor changing interactions are the decays $B^0_s \rightarrow \mu^\mp e^\pm$ and $B^+ \rightarrow K^+ \mu^\mp e^\pm$.
The existing experimental upper bounds do not impose any non-trivial constraint on the size of the $|B^e_{21}| = |B^e_{12}|$ coupling, and in general
we expect this branching fractions to be well below current limits (see Table~\ref{tab:BsDecays} and \ref{tab:BKdecays}):
\begin{equation}
	\mathcal{B} (B^0_s \rightarrow \mu^\mp e^\pm) \approx 1.6 \cdot 10^{-12} \left( \frac{3 \ {\rm TeV}}{1/R} \right)^4
		\left( \frac{|B^e_{21}|}{0.1} \right)^2	\left( \frac{|B^d_{L32}|}{0.5 |V^*_{tb} V_{ts}|} \right)^2
\end{equation}
\begin{equation}
	\mathcal{B} (B^+ \rightarrow K^+ \mu^\mp e^\pm) \approx  8.7 \cdot 10^{-11} \left( \frac{3 \ {\rm TeV}}{1/R} \right)^4
		\left( \frac{|B^e_{21}|}{0.1} \right)^2	\left( \frac{|B^d_{L32}|}{0.5 |V^*_{tb} V_{ts}|} \right)^2
\end{equation}

\section{Conclusions}
\label{sec:conclusions}

The very particular structure of the SM regarding flavor violation makes $b$-flavored meson decays a particularly
promising ground in the quest for NP.
Of particular interest is the $R_K \equiv \mathcal{B} (B^+ \rightarrow K^+ \mu^+ \mu^-) / \mathcal{B} (B^+ \rightarrow K^+ e^+ e^-)$ ratio,
which has been measured to differ by a factor of $\approx 0.75$ from its SM prediction.
Although the current significance of this discrepancy is only at the $2.6 \sigma$ level, future measurements could provide clear evidence for NP
if this deficit persists.
Moreover, other measurements concerning $b \rightarrow s$ transitions have also shown some level of disagreement with the SM,
in particular angular observables of the decay $B \rightarrow K^* \mu^+ \mu^-$~\cite{Aaij:2015oid},
and the branching fractions of the processes $B \rightarrow K^* \mu^+ \mu^-$~\cite{Aaij:2014pli} and $B^0_s \rightarrow \phi \mu^+ \mu^-$~\cite{Aaij:2013aln}.
Careful consideration of a comprehensive array of $b \rightarrow s$ data seems to prefer NP versus the SM alone~\cite{Altmannshofer:2014rta}.

In this work, we have studied these anomalies in the context of natural Scherk-Schwarz models~\cite{Dimopoulos:2014aua,Garcia:2015sfa} --
a class of theories that address the hierarchy problem combining SUSY with a flat extra dimension.
We have seen that $b \rightarrow s$ anomalies can be successfully accommodated in this context,
for a compactification scale $1/R = 3 - 4 \ {\rm TeV}$, which is the range preferred by naturalness.
The success of these models in accounting for the observed discrepancies in $b \rightarrow s$ physics is intrinsically linked
to the localization of the third generation on one of the 4D branes -- a structure motivated by naturalness considerations.
In fact, localizing only the third generation on one of the branes would be enough to accommodate the observed effects,
but strong experimental constraints on the decay $\tau \rightarrow 3 \mu$ rule out this possibility.
Instead, localizing the muon sector together with the third generation allows us to evade constraints from lepton flavor violating observables,
while retaining the success of accommodating $b \rightarrow s$ anomalies.

We have seen that, in the region of parameter space preferred by $R_K$ and other $b \rightarrow s$ measurements,
other observables are also sharply predicted to deviate from its SM values.
Of particular interest is the decay $B^0_s \rightarrow \mu^+ \mu^-$, whose branching ratio has been measured to be smaller than its SM prediction
by a factor $\approx 0.77$.
In the natural Scherk-Schwarz models we discuss, this supression factor is predicted to be similar to $R_K \approx 0.75$ --
certainly in agreement with observations.
More precise measurements of $\mathcal{B}(B^0_s \rightarrow \mu^+ \mu^-)$ will provide a key test of these models.
Similarly, given the structure of these theories, branching fractions of the decays  $B^0_s \rightarrow \tau^+ \tau^-$ and $B^+ \rightarrow K^+ \tau^+ \tau^-$
are also predicted to deviate from their SM values by a factor similar to $R_K$ -- measuring these observables in the future
will also be of great importance to assess the validity of these scenarios.

Finally, we have discussed how, in the lepton sector, the flavor structure of these theories is most constrained
by the decays $\tau^- \rightarrow e^- \mu \mu$, $\tau \rightarrow 3\mu$, and $\mu \rightarrow 3 e$.
Current experimental upper bounds on these three decays impose non-trivial constraints on lepton flavor violating couplings,
and imply that leptonic and semi-leptonic $B$ meson decays of the form $B^0_s \rightarrow l^\pm l^{' \mp}$ and $B^+ \rightarrow K^+ l^\pm l^{' \mp}$
(with $l' \neq l$)
are predicted to be several orders of magnitude below current experimental limits.


\section*{Acknowledgements}

We are very thankful to Uli Haisch and John March-Russell for useful discussions.
IGG is financially supported by the STFC and a Scatcherd European Scholarship from the University of Oxford.


\appendix

\section{$B^0_s - \overline{B}^0_s$ Oscillations}
\label{app:Bsmixing}

Oscillation phenomena in the $B^0_s - \overline{B}^0_s$ meson system is well described by the off-diagonal elements of the mass matrix $M^s_{12}$ and
the decay matrix $\Gamma^s_{12}$, and the relative phase between both $\phi_s =\arg(-M^s_{12}/\Gamma^s_{12})$.
The meaningful physical observables to be defined from these three quantities are the mass difference between mass-eigenstates $\Delta M_s$,
the decay rate difference $\Delta \Gamma_s$, and the CP asymmetry $a^s_{sl}$.
These are given by
\begin{equation}
	 \Delta M_s = 2 | M_{12}^s |	\ ,	\qquad	\Delta \Gamma_s = 2 | \Gamma_{12}^s | \cos \phi_s \ ,	\qquad	a^s_{sl} = \left| \frac{\Gamma_{12}^s}{M_{12}^s} \right| \sin \phi_s \ .
\label{eq:BsObservables}
\end{equation}

On the other hand, the set of effective operators relevant for the description of $B^0_s - \overline{B}^0_s$ mixing are
\begin{equation}
\begin{aligned}
	\mathcal{O}_1^s &= (\bar b_L^i \gamma^\mu s_L^i) (\bar b_L^j \gamma_\mu s_L^j)&
		\qquad {\mathcal{O}}_1^{\prime s} &= \mathcal{O}_1^s (L \leftrightarrow R)\\
	\mathcal{O}_2^s &= (\bar b_R^i s_L^i) (\bar b_R^j s_L^j)&
		\qquad {\mathcal{O}}_2^{\prime s} &= \mathcal{O}_2^s (L \leftrightarrow R)\\
	\mathcal{O}_3^s &= (\bar b_R^i s_L^j) (\bar b_R^j s_L^i)&
		\qquad {\mathcal{O}}_3^{\prime s} &= \mathcal{O}_3^s (L \leftrightarrow R)\\
	\mathcal{O}_4^s &=(\bar b_L^i s_R^i) (\bar b_R^j s_L^j)&
		\qquad \mathcal{O}_5^s &=(\bar b_L^i s_R^j) (\bar b_R^j s_L^i) \ ,
\end{aligned}
\label{eq:BsOperators}
\end{equation}
and non-zero contributions to their Wilson coefficients arise in natural Scherk-Schwarz theories after integrating out
the KK-modes of the neutral gauge bosons ($Z$, photon, and gluons, with the latter providing the dominant effect).
These NP contributions read, at scale $1/R$:
\begin{equation}
\begin{aligned}
	& C_{1, {\rm NP}}^s \simeq \frac{\pi^2}{6} \frac{1}{(1/R)^2} (B^d_{L32})^2
		\left( \frac{1}{3} g_s^2 + \frac{e^2}{9} + \frac{g^2}{c_W^2} (\alpha^d_L)^2 \right) \\
	& C_{1, {\rm NP}}^{\prime s} \simeq \frac{\pi^2}{6} \frac{1}{(1/R)^2} (B^d_{R32})^2
		\left( \frac{1}{3} g_s^2 + \frac{e^2}{9} + \frac{g^2}{c_W^2} (\alpha^d_R)^2 \right) \\
	& C_{2, {\rm NP}}^s = C_{2, {\rm NP}}^{\prime s} = 0 \\ 
	& C_{3, {\rm NP}}^s = C_{3, {\rm NP}}^{\prime s} = 0 \\ 
	& C_{4, {\rm NP}}^s \simeq \frac{\pi^2}{6} \frac{1}{(1/R)^2} (B^d_{L32} B^d_{R32}) (-2 g_s^2) \\
	& C_{5, {\rm NP}}^s \simeq \frac{\pi^2}{6} \frac{1}{(1/R)^2} (B^d_{L32} B^d_{R32})
		\left( \frac{2}{3} g_s^2 - \frac{4 e^2}{9}  - \frac{4 g^2}{c_W^2} \alpha^d_L \alpha^d_R\right).
\end{aligned}
\label{eq:BsWilsonCoeff}
\end{equation}

The off-diagonal matrix element $M^s_{12}$ is then given by
\begin{equation}
	M^s_{12} = \frac{1}{2 m_{B_s}} {\langle \overline{B}_s^0 | \mathcal{H}_{eff} | B_s^0 \rangle}^* \ ,
\end{equation}
where
\begin{equation}
	\langle \overline{B}_s^0 | \mathcal{H}_{eff} | B_s^0 \rangle = \sum_i C_i^s (\mu_b) \langle \overline{B}_s^0 | \mathcal{O}_i^s (\mu_b) | B_s^0 \rangle \ .
\end{equation}
Here, $\langle \overline{B}_s^0 | \mathcal{O}_i^s (\mu_b) | B_s^0 \rangle$ refers to the matrix elements of the effective operators (evaluated in the lattice at scale $\mu_b = 4.6 \ {\rm GeV}$),
and $C_i^s (\mu_b)$ are the values of the Wilson coefficients once they have been RG-evolved down to scale $\mu_b$.
The hadronic matrix elements can be written as
\begin{equation}
\begin{aligned}
	\langle \overline{B}_s^0 | \mathcal{O}_1^s (\mu_b) | B_s^0 \rangle & = \langle \overline{B}_s^0 | {\mathcal{O}}_1^{\prime s} (\mu_b) | B_s^0 \rangle
					= \frac{2}{3} m_{B_s}^2 f_{B_s}^2 B_1^s(\mu_b) \\
	\langle \overline{B}_s^0 | \mathcal{O}_4^s (\mu_b) | B_s^0 \rangle & = \frac{1}{2} \left\{ \frac{m_{B_s}}{m_b(\mu_b) + m_s(\mu_b)} \right\}^2 m_{B_s}^2 f_{B_s}^2 B_4^s(\mu_b) \\
	\langle \overline{B}_s^0 | \mathcal{O}_5^s (\mu_b) | B_s^0 \rangle & = \frac{1}{6} \left\{ \frac{m_{B_s}}{m_b(\mu_b) + m_s(\mu_b)} \right\}^2 m_{B_s}^2 f_{B_s}^2 B_5^s(\mu_b) \ ,
\end{aligned}
\end{equation}
and we take $m_{B_s} = 5.367 \ {\rm GeV}$~\cite{Olive:2016xmw}, $f_{B_s} = 0.235 \ {\rm GeV}$~\cite{Christ:2014uea},
and $B^s_1 = 0.85$, $B^s_4 = 1.10$, $B^s_5 = 2.02$~\cite{Carrasco:2013zta}.
We evolve the Wilson coefficients at scale $1/R$ down to scale $\mu_b$ using the appropriate RG evolution equations~\cite{Becirevic:2001jj},
and note that RG evolution does not generate a non-zero contribution to $C_{2, {\rm NP}}^{(\prime) s}$ or $C_{3, {\rm NP}}^{(\prime) s}$.

Finally, we find it is convenient to write $M_{12}^s$ and $\Gamma_{12}^s$ as follows:
\begin{equation}
	M_{12}^s \equiv M_{12, {\rm SM}}^s | r_s | e^{i \Delta \phi_s}		\qquad	{\rm and} \qquad
			\Gamma_{12}^s \equiv \Gamma_{12, {\rm SM}}^s | r'_s | e^{i \Delta \phi'_s} \ ,
\end{equation}
so that $|r^{(')}_s|$ and $\Delta \phi_s^{(')}$ parametrize the effect of NP on the magnitude and phase of  $M_{12}^s$ ($\Gamma_{12}^s$) compared to their SM values.
This allows for the observables in Eq.(\ref{eq:BsObservables}) to be conveniently written as
\begin{equation}
	\frac{\Delta M_s}{\Delta M_{s, {\rm SM}}} = |r_s| \ ,		\qquad
			 \frac{\Delta \Gamma_s}{\Delta \Gamma_{s, {\rm SM}}} = |r'_s| \frac{\cos \phi_s}{\cos \phi_{s, {\rm SM}}}	\ ,	\qquad
			\frac{a^s_{sl}}{a^s_{sl, {\rm SM}}} = \frac{|r'_s|}{|r_s|} \frac{\sin \phi_s}{\sin \phi_{s, {\rm SM}}} \ ,
\end{equation}
where $\phi_s = \phi_{s, {\rm SM}} + \Delta \phi_s - \Delta \phi'_s$, and notice that it may be written as $\phi_s = \tan^{-1}(a^s_{sl} \Delta M_s / \Delta \Gamma_s)$.
In the kind of scenarios considered in this work, no extra contributions to $\Gamma_{12}$ arise, and so $| r'_s |=1$ and $\Delta \phi'_s = 0$.
On the other hand, non-zero $|B^d_{L32}|$ and $\delta$ (remember our parametrization of $B^d_{L32}$ in Eq.(\ref{eq:BdL32})) lead to $|r_s| \neq 1$ and $\Delta \phi_s \neq 0$,
which may be conveniently written as
\begin{equation}
	|r_s| = \sqrt{ 1 + 2 \epsilon_s \cos 2 \delta + \epsilon_s^2 } \ ,
\end{equation}
\begin{equation}
	\cos \Delta \phi_s = \frac{1 + \epsilon_s \cos 2\delta}{|r_s|} \qquad {\rm and} \qquad
	\sin \Delta \phi_s = - \frac{\epsilon_s \sin 2\delta}{|r_s|}
\end{equation}
where
\begin{equation}
	\epsilon_s \equiv \frac{ | M_{12, {\rm NP}}^s | }{ | M_{12, {\rm SM}}^s | } = \frac{ \Delta M_{s, {\rm NP}} }{ \Delta M_{s, {\rm SM}} } \ .
\end{equation}

Then, computing $\Delta M_{s, {\rm NP}}$ as described in this section, and taking into account the experimental measurements and SM predictions
of $B^0_s - \overline{B}^0_s$ oscillation observables summarized in Table~\ref{tab:BsParameters}, we can extract constraints on the size of $|B^d_{L32}|$ and $\delta$,
as shown in Fig.~\ref{fig:contour}.
\begin{table}[h]
  \begin{center}
    \begin{tabular}{ l | l | l }
	 & Experimental measurement & SM prediction\\
	\hline
	\hline
	$\Delta M_s$		&	$17.757 \pm 0.021 \ {\rm ps}^{-1}$ \cite{Amhis:2014hma}		&		$(18.3 \pm 2.7) \ {\rm ps}^{-1}$ \cite{Artuso:2015swg}\\
	$\Delta \Gamma_s$	&	$0.081 \pm 0.006 \ {\rm ps}^{-1}$ \cite{Amhis:2014hma}		&		$(0.085 \pm 0.015) \ {\rm ps}^{-1}$ \cite{Artuso:2015swg}\\
	$a^s_{sl}$			&	$(17 \pm 30) \cdot 10^{-4}$ \cite{Aaij:2016yze}			&		$(0.222 \pm 0.027) \cdot 10^{-4}$ \cite{Artuso:2015swg}\\
	\hline
	$\phi_s$			&	$ [-0.30, 0.84] $										&		$0.0048 \pm 0.0012$
    \end{tabular}
    \caption{Experimental measurement and SM prediction for the three observables relevant to oscillation phenomena in the $B_s^0$ sector:
		$\Delta M_s$, $\Delta \Gamma_s$, and $a^s_{sl}$.
		The last line features the experimental constraint on and SM prediction of $\phi_s$, obtained through the relation
		$\phi_s = \tan^{-1}(a^s_{sl} \Delta M_s / \Delta \Gamma_s)$ from the other quantities quoted in the table.}
    \label{tab:BsParameters}  
  \end{center}
\end{table}

\section{(Semi-)Leptonic $B$ Decays}

\subsection*{$B^0_s$ leptonic decays}
\label{app:Bsmumu}

The relevant set of effective operators for the study of the decays $B^0_s \rightarrow l^- l^{\prime +}$ (allowing for $l^\prime \neq l$)
are $\mathcal{O}^{(')}_9$ and $\mathcal{O}^{(')}_{10}$ in Eq.(\ref{eq:operatorsRK}), generalized to include different lepton flavors, i.e.
\begin{equation}
\begin{aligned}
	\mathcal{O}^{ll'}_9 &= (\bar b \gamma^\nu P_L s) (\bar l \gamma_\nu l') & \qquad \mathcal{O'}^{ll'}_9 &= (\bar b \gamma^\nu P_R s) (\bar l \gamma_\nu l') \\
	\mathcal{O}^{ll'}_{10} &= (\bar b \gamma^\nu P_L s) (\bar l \gamma_\nu \gamma^5 l') & \qquad \mathcal{O'}^{ll'}_{10} &= (\bar b \gamma^\nu P_R s) (\bar l \gamma_\nu \gamma^5 l')
\label{eq:operatorsBs2leptons}
\end{aligned}
\end{equation}

In the lepton flavor preserving case only $\mathcal{O}^{ll}_{10}$ and $\mathcal{O'}^{ll}_{10}$ are relevant,
and the ratio of the prediction for $\mathcal{B} (B^0_s \rightarrow l^- l^{+})$ to its SM value can be written as
\begin{equation}
	R^s_{l l} \equiv \frac{\mathcal{B} (B_s \rightarrow l^+ l^-)}{\mathcal{B} (B_s \rightarrow l^+ l^-)_{\rm SM}} =
		\frac{| C_{10, {\rm SM}} + C^{l l}_{10, {\rm NP}} - C^{\prime l l}_{10, {\rm NP}} |^2}{ |C_{10, {\rm SM}}|^2 } \ ,
\end{equation}
where the NP Wilson coefficients are those of Eq.(\ref{eq:WilsonCoeff}), in general
\begin{equation}
\begin{aligned}
	C^{l l}_{10, {\rm NP}} & \simeq - \frac{8 \pi^4}{3} \frac{v^2}{(1/R)^2} \frac{B^d_{L32}}{V^*_{tb} V_{ts}} ( - S_{LL} B^e_{Lll} + S_{LR} B^e_{Rll} ) \\
	C^{' l l}_{10, {\rm NP}} & \simeq - \frac{8 \pi^4}{3} \frac{v^2}{(1/R)^2} \frac{B^d_{R32}}{V^*_{tb} V_{ts}} ( - S_{RL} B^e_{Lll} + S_{RR} B^e_{Rll} ) \ ,
\end{aligned}
\end{equation}
with $B^e_{Jee} = B^e_{J11}$, etc.

In Table~\ref{tab:BsDecays}, we summarize the current measurements and upper bounds for $B^0_s \rightarrow l^+ l^-$ decays, together with their SM predictions.
Of particular interest is $B^0_s \rightarrow \mu^+ \mu^-$, the only leptonic decay for which an actual measurement exists.
The SM prediction for this branching fraction is $\mathcal{B} (B_s \rightarrow \mu^+ \mu^-)_{\rm SM} = (3.65 \pm 0.23) \cdot 10^{-9}$~\cite{Bobeth:2013uxa},
and the experimental measurement $\mathcal{B} (B_s \rightarrow \mu^+ \mu^-) = 2.8^{+ 0.7}_{-0.6} \cdot 10^{-9}$~\cite{CMS:2014xfa},
leading to a ratio $R^s_{\mu \mu} = 0.77^{+0.20}_{-0.17}$.
Although the uncertainty in this ratio is somewhat large, values smaller than unity seem to be preferred.

On the other hand, in the lepton flavor violating case, all four operators in Eq.(\ref{eq:operatorsBs2leptons}) are relevant,
and the branching fraction for the decay $B^0_s \rightarrow l^\mp l^{\prime \pm}$ can be written as
\begin{equation}
	\mathcal{B} (B^0_s \rightarrow l^\mp {l'}^\pm) = 2 \ \frac{\Gamma (B^0_s \rightarrow l^- {l'}^+) }{\Gamma_{B_s}} \ ,
\end{equation}
where
\begin{equation}
\begin{aligned}
	\Gamma (B^0_s \rightarrow l^- {l'}^+) \simeq & \frac{\alpha^2 | V_{tb}^* V_{ts} |^2}{64 \pi^3 v^4} f_{B_s}^2 \frac{m_{B_s}^2 - m_l^2}{m_{B_s}} \times \\
				& (E_q E_{q'} - E_q^2 + m_l^2) \left\{ | C^{l l'}_{9, {\rm NP}} - C^{\prime l l'}_{9, {\rm NP}} |^2 + | C^{l l'}_{10, {\rm NP}} - C^{\prime l l'}_{10, {\rm NP}} |^2 \right\} \ ,
\end{aligned}
\end{equation}
with $E_q = (m_{B_s}^2 + m_l^2) / (2 m_{B_s})$, $E_{q'}= \sqrt{E_q^2 - m_l^2}$, and this expression is valid for $m_l \gg m_{l'}$.
We take $m_{B_s} = 5.367 \ {\rm GeV}$~\cite{Olive:2016xmw}, $f_{B_s} = 0.235 \ {\rm GeV}$~\cite{Christ:2014uea},
and $\Gamma_{B_s}^{-1}=1.505 \ {\rm ps}$~\cite{Amhis:2014hma}.
The relevant Wilson coefficients are those of Eq.(\ref{eq:WilsonCoeff}) after allowing for $l' \neq l$ (with $B^e_{J \tau \mu} = B^e_{J 3 2}$, etc.).
The only process of this kind with a reported upper bound is the $B_s^0 \rightarrow \mu^\pm e^\mp$ decay, as shown in Table~\ref{tab:BsDecays}.

\begin{table}[h]
  \begin{center}
    \begin{tabular}{ l | l | l }
	 				& Experimental measurement & SM prediction\\
	\hline
	\hline
	$\mathcal{B}(B^0_s \rightarrow e^+ e^-)$			&	$< 2.8 \cdot 10^{-7}$~\cite{Olive:2016xmw}		 	&	$(8.54 \pm 0.55) \cdot 10^{-14}$~\cite{Bobeth:2013uxa}\\
	$\mathcal{B}(B^0_s \rightarrow \mu^+ \mu^-)$		&	$2.8^{+ 0.7}_{-0.6} \cdot 10^{-9}$~\cite{CMS:2014xfa}	&	$(3.65 \pm 0.23) \cdot 10^{-9}$~\cite{Bobeth:2013uxa}\\
	$\mathcal{B}(B^0_s \rightarrow \tau^+ \tau^-)$		&	$< 3.0 \cdot 10^{-3}$~\cite{BstautauExp} 			&	$(7.73 \pm 0.49) \cdot 10^{-7}$~\cite{Bobeth:2013uxa}\\
	\hline
	$\mathcal{B}(B^0_s \rightarrow \tau^\pm \mu^\mp)$		&	no bound reported				&  \\
	$\mathcal{B}(B^0_s \rightarrow \tau^\pm e^\mp)$		&	no bound reported				&  \\
	$\mathcal{B}(B^0_s \rightarrow \mu^\pm e^\mp)$		&	$< 1.4 \cdot 10^{-8}$~\cite{Aaij:2013cby}	& 
    \end{tabular}
    \caption{Experimental measurement (or upper bounds) for the branching fractions of different leptonic decays channels of the $B^0_s$ meson, together with their predictions in the SM.
	The only measured branching fraction is that corresponding to the decay $B_s \rightarrow \mu^+ \mu^-$, which falls bellow the SM value.
	SM predictions for lepton flavor violating decays are negligible.}
    \label{tab:BsDecays}  
  \end{center}
\end{table}

\subsection*{$B^+$ semi-leptonic decays}
\label{app:BtoK2leptons}

The relevant set of effective operators relevant for the study of $B^+ \rightarrow K^+ l^- l^{'+}$ decays in this context are the same as those in Eq.(\ref{eq:operatorsBs2leptons}).
As mentioned in Sec.~\ref{sec:RKformalism}, a good approximation for the branching fraction of the lepton flavor preserving decay $B^+ \rightarrow K^+ l^+ l^{-}$ is
\begin{equation}
\begin{aligned}
	\mathcal{B} (B^+ \rightarrow K^+ l^+ l^{-}) \simeq & \ \mathcal{B} (B^+ \rightarrow K^+ l^+ l^{-})_{\rm SM} \ \times \\
		& \ \ \frac{| C_{9, {\rm SM}} + C^{ll}_{9, {\rm NP}} + C^{' ll}_{9, {\rm NP}} |^2 + | C_{10, {\rm SM}} + C^{ll}_{10, {\rm NP}} + C^{' ll}_{10, {\rm NP}} |^2}{|C_{9, {\rm SM}}|^2 + |C_{10, {\rm SM}}|^2} \ .
\end{aligned}
\label{eq:BKll}
\end{equation}

Regarding the flavor violating decay $B^+ \rightarrow K^+ l^\mp {l'}^\pm$, its branching ratio can be approximated as
\begin{equation}
\begin{aligned}
	\mathcal{B} (B^+ \rightarrow K^+ l^\mp {l'}^\pm) \simeq & \ 2 \ \mathcal{B} (B^+ \rightarrow K^+ l^+ l^{-})_{\rm SM} \ \times \\
		& \ \ \frac{| C^{ll'}_{9, {\rm NP}} + C^{' ll'}_{9, {\rm NP}} |^2 + | C^{ll'}_{10, {\rm NP}} + C^{' ll'}_{10, {\rm NP}} |^2}{|C_{9, {\rm SM}}|^2 + |C_{10, {\rm SM}}|^2} \ ,
\end{aligned}
\label{eq:BKllp}
\end{equation}
an expression that is approximately true in the case where $m_{l'} \approx m_l$, which is clearly not the case for $l=\tau$ and $l' = e, \mu$.
Taking into account the actual masses of the final leptons will only make an $\mathcal{O}(1)$ difference to the final result --
this would not affect our conclusions, since the maximum possible values predicted for these branching fractions are well below currents experimental upper bounds,
as discussed in Sec.~\ref{sec:FlavorViolatingDecays}.

Table~\ref{tab:BKdecays} shows all measurements (or upper bounds) and SM predictions for this type of semileptonic decays.

\begin{table}[h]
  \begin{center}
    \begin{tabular}{ l | l | l }
	 				& Experimental measurement & SM prediction\\
	\hline
	\hline
	$\mathcal{B}(B^+ \rightarrow K^+ e^+ e^-)$ \ (*)		&	$1.56^{+0.19 \ +0.06}_{-0.15 \ -0.04} ) \cdot 10^{-7}$\cite{Aaij:2014ora}		&	$1.75^{+0.60}_{-0.29} \cdot 10^{-7}$\cite{Bobeth:2012vn}\\
	$\mathcal{B}(B^+ \rightarrow K^+ \mu^+ \mu^-)$ (*)	&	$1.19 \pm 0.03 \pm 0.06) \cdot 10^{-7}$\cite{Aaij:2014pli}				&	$1.75^{+0.60}_{-0.29} \cdot 10^{-7}$\cite{Bobeth:2012vn}\\
	$\mathcal{B}(B^+ \rightarrow K^+ \tau^+ \tau^-)$	&	$< 2.25 \cdot 10^{-3}$\cite{TheBaBar:2016xwe}						&	$(1 - 2) \cdot 10^{-7}$\cite{Bouchard:2013mia,Hewett:1995dk} \\
	\hline
	$\mathcal{B}(B^+ \rightarrow K^+ \tau^\pm \mu^\mp)$		&$<4.8 \cdot 10^{-5}$~\cite{Olive:2016xmw}	\\
	$\mathcal{B}(B^+ \rightarrow K^+ \tau^\pm e^\mp)$		&	$<3.0 \cdot 10^{-5}$~\cite{Olive:2016xmw}	\\
	$\mathcal{B}(B^+ \rightarrow K^+ \mu^\pm e^\mp)$		&	$<9.1 \cdot 10^{-8}$~\cite{Olive:2016xmw}	
	
    \end{tabular}
    \caption{Experimental measurement (or upper bounds) for the branching fractions of different semi-leptonic decay channels of the $B^+$ meson,
	together with their predictions in the SM.
	(*): measurement and theoretical prediction in the dilepton invariant mass region $1<q^2 / {\rm GeV}^2<6$.
	SM predictions for lepton flavor violating decays are negligible.}
    \label{tab:BKdecays}  
  \end{center}
\end{table}

\section{Lepton Decays}
\label{app:LeptonDecays}

The set of effective operators relevant for the description of leptonic decays of the form $l_1^- \rightarrow l_2^- l^+ l^-$ are
\begin{equation}
\begin{aligned}
	\mathcal{O}^{l_1 l_2}_{\rm VV} &= (\bar l_2 \gamma^\mu l_1) (\bar l \gamma_\mu l) &
			\qquad \mathcal{O}^{l_1 l_2}_{\rm AA} &= (\bar l_2 \gamma^\mu \gamma^5 l_1) (\bar l \gamma_\mu \gamma^5 l) \\
	\mathcal{O}^{l_1 l_2}_{\rm VA} &= (\bar l_2 \gamma^\mu l_1) (\bar l \gamma_\mu \gamma^5 l) &
			\qquad \mathcal{O}^{l_1 l_2}_{\rm AV} &= (\bar l_2 \gamma^\mu \gamma^5 l_1) (\bar l \gamma_\mu l)
\label{eq:leptondecay}
\end{aligned}
\end{equation}

In the context of natural Scherk-Schwarz models as those described in Sec.~\ref{sec:structure}, the branching fractions for the decays $\mu \rightarrow 3e$,
$\tau \rightarrow 3 \mu$, and $\tau^- \rightarrow e^- \mu \mu$ are given by
\begin{equation}
\begin{aligned}
	\mathcal{B} (\mu \rightarrow 3e) & \simeq \frac{\Gamma (\mu \rightarrow 3e) }{\Gamma (\mu \rightarrow e \nu_\mu \bar \nu_e) } \\
			& \simeq \left( \frac{\pi^2}{6} \right)^2 \frac{v^4}{(1/R)^4} \times \\
			& \ \ (2 A_{LL} |B^e_{L11} B^e_{L12}|^2 + 2 A_{RR} |B^e_{R11} B^e_{R12}|^2 + A_{LR} |B^e_{L11} B^e_{R12}|^2 + A_{RL} |B^e_{R11} B^e_{L12}|^2) \ ,
\end{aligned}
\end{equation}
\begin{equation}
\begin{aligned}
	\mathcal{B} (\tau \rightarrow 3 \mu)
			& \simeq \frac{1}{\Gamma_\tau} \frac{m_\tau^5}{2^7 3 \pi^3} \left( \frac{\pi^2}{6} \right)^2 \frac{1}{(1/R)^4} \times \\
			& \ \ (2 A_{LL} |B^e_{L22} B^e_{L23}|^2 + 2 A_{RR} |B^e_{R22} B^e_{R23}|^2 + A_{LR} |B^e_{L22} B^e_{R23}|^2 + A_{RL} |B^e_{R22} B^e_{L23}|^2) \ ,
\end{aligned}
\end{equation}
and
\begin{equation}
\begin{aligned}
	\mathcal{B} (\tau \rightarrow e \mu \mu)
			& \simeq \frac{1}{\Gamma_\tau} \frac{m_\tau^5}{2^8 3 \pi^3} \left( \frac{\pi^2}{6} \right)^2 \frac{1}{(1/R)^4} \times \\
			& \ \ (A_{LL} |B^e_{L22} B^e_{L13}|^2 + A_{RR} |B^e_{R22} B^e_{R13}|^2 + A_{LR} |B^e_{L22} B^e_{R13}|^2 + A_{RL} |B^e_{R22} B^e_{L13}|^2) \ ,
\end{aligned}
\end{equation}
where $A_{LL} = (g^2/\cos^2 \theta_w) {\alpha^{e}_L}^2  + (e Q_e)^2$, $A_{RR} = (g^2/\cos^2 \theta_w) {\alpha^e_R}^2 +  (e Q_e)^2$, $A_{LR} = (g^2/\cos^2 \theta_w) \alpha^e_L \alpha^e_R + (e Q_e)^2$,
and $A_{RL} = A_{LR}$.
We have used $m_\tau = 1.78 \ {\rm GeV}$, $\Gamma_\tau = 290.3 \cdot 10^{-15}$~\cite{Olive:2016xmw}.

\bibliography{flavourReferences}

\end{document}